\documentclass[aip,reprint]{revtex4-1}
\usepackage{graphicx, float}
\draft
\begin{document}

\title{Dual Mode Operation of a  Hydromagnetic Plasma Thruster to Achieve Tunable Thrust and Specific Impulse} 

\author{Thomas C. Underwood}
\affiliation{Department of Aerospace Engineering and Engineering Mechanics, University of Texas at Austin, TX, USA}
\email{Author correspondence should be addressed to: thomas.underwood@utexas.edu}
\author{William M. Riedel}
\affiliation{Stanford Plasma Physics Laboratory, Mechanical Engineering Department, Stanford University, Stanford, CA, USA}
\author{Mark A. Cappelli}
\affiliation{Stanford Plasma Physics Laboratory, Mechanical Engineering Department, Stanford University, Stanford, CA, USA}

\date{\today}

\begin{abstract}
We report here on initial studies of a pulsed hydromagnetic plasma gun that can operate in either a pre-filled or a gas-puff mode on demand. These modes enable agile and responsive performance through tunable thrust and specific impulse. Operation with a molecular nitrogen propellant is demonstrated to show that the hydromagnetic thruster is a candidate technology for air-harvesting and drag compensation in very low Earth orbit. Dual mode operation is achieved by leveraging propellant gas dynamics to change the fill fraction and flow collisionality within the thruster. This results in the formation of distinct modes that are characterized by the current-driven hydromagnetic waves that they allow, namely a magneto-deflagration and magneto-detonation respectively. These modes can be chosen by changing the time propellant is allowed to diffuse into the thruster based on the desired performance. Using time-of-flight emission diagnostics to characterize near-field flow velocities, we find that a relatively dramatic transition occurs between modes, with exhaust velocities ranging from 10 km/s to 55 km/s in deflagration and detonation regimes, respectively. Simulations of the processed mass bit offers a first glimpse into possible thruster performance confirming a broad range and tradeoff between specific impulse (2600 - 5600 sec) and thrust (up to 31 mN) when operating in a burst mode.
\end{abstract}


\maketitle


\section{Introduction}

Propulsion technologies continue to be limited by their inability to offer agile and responsive performance. A common example is the tradeoff between the efficiency of propellant usage (specific impulse, I$_{\text{sp}}$) and the generation of thrust (T) \cite{peukert2014ohb}. In systems ranging from chemical rockets to electric thrusters, any increase in thrust generation is typically accompanied with a decrease in the acceleration efficiency of propellant. This can constrain the scope of missions and, in some cases, require the use of multiple propulsion systems to satisfy the required altitude, duration, weight, and level of thrust generation. Increasingly, there is a need for emerging technologies that can rapidly change their operational performance (i.e., T or I$_{\text{sp}}$) in response to changing mission objectives. Added performance agility can allow a single thruster to smoothly transition from optimizing propellant usage to thrust generation while in orbit.

Electric propulsion (EP) in particular uniquely leverages electric power (P) and magnetic fields to accelerate propellant \cite{mazouffre2016electric}. This allows EP technology to operate efficiently in applications that demand thrust generation and I$_{\text{sp}}$ levels spanning many orders of magnitude. The use of external electrical power also adds a level of control over the operational characteristics of the thruster. For example, depending on the manner in which energy is supplied, EP technology can be categorized as either an electrospray or a plasma-based electrostatic, electromagnetic, or electrothermal thruster. Research into these systems has focused on uncovering mechanisms that enable reliable and sustained operation at either high thrust efficiency (T/P) \textit{or} I$_{\text{sp}}$. Little work however has been done to explore reconfigurability in EP systems and how they might allow rapid and on-demand transitions between values of high T/P \textit{and} I$_{\text{sp}}$.

Beyond reconfigurability, simply operating EP systems in different gas dynamic environments is a challenge that has been largely unexplored. Recent efforts \cite{cifali2011preliminary} to use plasma thrusters in an air breathing configuration for very low Earth orbit (VLEO) conditions has been met with significant performance decreases. Measurements of the plume composition in a low power Z-70 (500W) Hall effect thruster (HET) operating on Xe/Air mixtures confirms the presence of ions of both molecular and atomic states suggesting that a considerable amount of energy may be invested in dissociation \cite{gurciullo2019ion}. In recent unpublished studies in our laboratory, the transition from Xe (100\%) to Xe-N$_{2}$ (17\%-83\%) mixtures in the same Z-70  thruster reduced the thrust from 24 mN to 13 mN and the I$_{\text{sp}}$ from 1270 s to 820 s for comparable thruster power. It is still uncertain if electrothermal or inductive thruster technology can generate the I$_{\text{sp}}$ or thrust requirements for sustained VLEO operation \cite{dietz2019molecular, romano2018system, romano2020rf}. Pulsed plasma thrusters (PPTs) on the other hand have shown the ability to still operate efficiently on molecular propellants such as water vapor \cite{ziemer2002performance} or N$_{2}$ \cite{larson1965energy} due in part to their higher input energies. Experiments have also found that differing levels of propellant loading within PPTs results in distinct operational characteristics \cite{mather1964investigation, loebner2015evidence}. These observations motivate the exploration of gas dynamic loading in PPTs as a controllable means to enable reconfigurability and agile operation.

\begin{figure*}[tp]
\begin{center}
\includegraphics[width=1\textwidth]{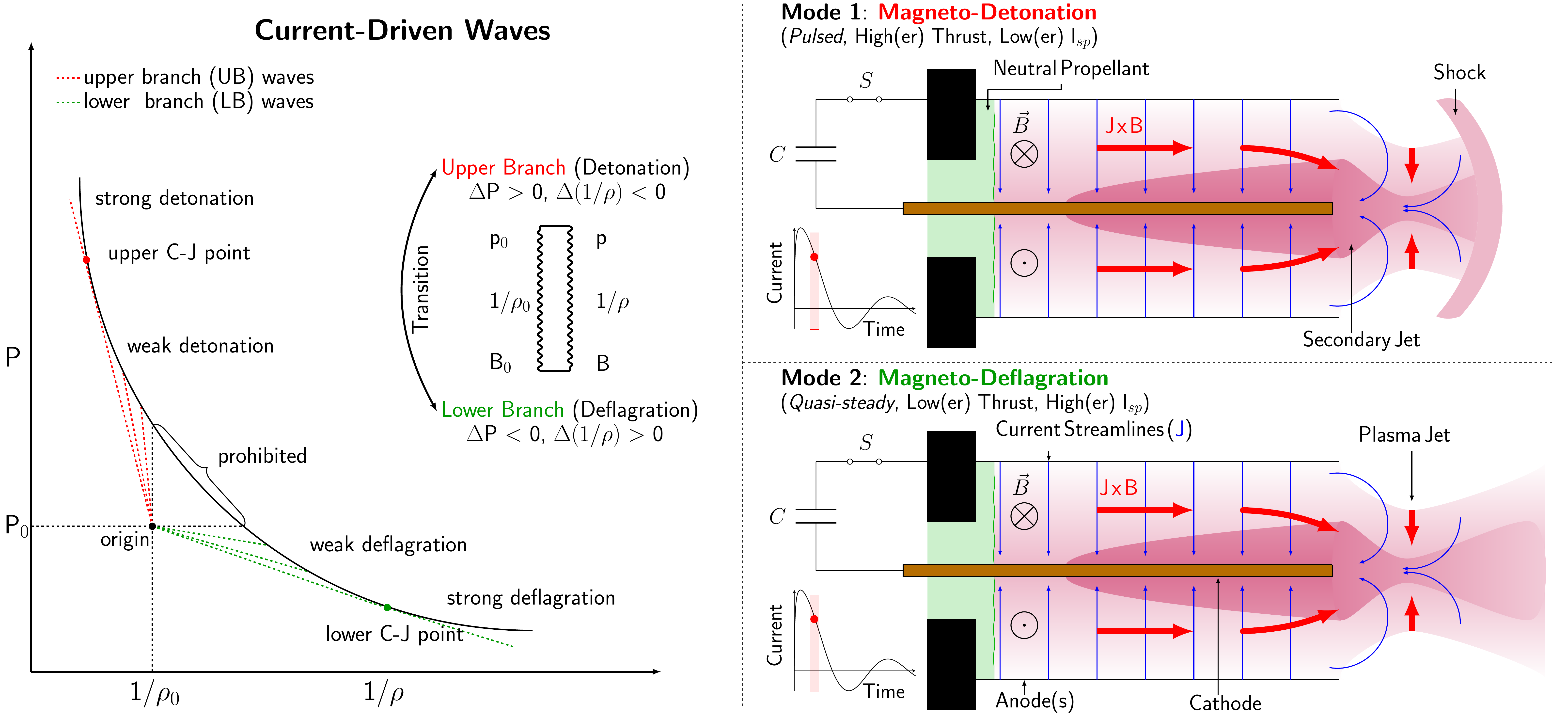}
\end{center}
\caption{Schematic depicting the operational modes of a hydromagnetic plasma thruster. The first mode, called a magneto-detonation wave, generates a transient shock that produces high(er) thrust and low(er) I$_{\text{sp}}$. The second mode, called a magneto-deflagration wave, generates a quasi-steady expansion wave that produces low(er) thrust and high(er) I$_{\text{sp}}$. The transition between these operational modes can be controlled by initializing the gas dynamic state of the thruster.}\label{fig:1}
\end{figure*}

This paper describes a novel pulsed hydromagnetic thruster that can operate in two distinct modes on demand. The first, called a magneto-detonation, is a transient mode that enables the selective generation of higher thrust levels. The magneto-detonation is characteristic of the upper branch of a current-driven ionization wave \cite{loebner2015evidence}. The second, called a magneto-deflagration, is a quasi-stationary mode that enables high exhaust velocities and thus I$_{\text{sp}}$. The magneto-deflagration is characterized by the lower branch of the wave. We review the theory and present an experimental characterization of these modes while the thruster is operating on N$_{2}$ propellant. Finally, we show how propellant loading and flow collisionality can be used to enable reliable mode selection within the thruster. 


\section{Dual Mode Plasma Thruster}

\subsection{Facility Description}

The Stanford hydromagnetic thruster operates by ionizing and accelerating propellant using an induced Lorentz force. The device, along with other pulsed Lorentz accelerators, is an extension of the original Marshall plasma gun \cite{marshall1960performance}. Since their inception, Marshall guns have been used for applications ranging from simulated astrophysics \cite{underwood2017plasma}, plasma jet driven magneto-inertial fusion (PJMIF) \cite{thio2019plasma}, z-pinch schemes \cite{shumlak2001evidence}, material ablation \cite{UNDERWOOD201997}, and even neutron production \cite{zhang2019sustained}. Despite studies in the 1960's on the potential use of these hydromagnetic plasma guns in space propulsion \cite{ziemer2000review}, there has been little effort since the dawn of diverse electric propulsion demonstrations (1980's) to develop them as a source of thrust for space propulsion.

What makes hydromagnetic plasma guns unique among thrusters is that they are designed to operate with much higher voltages ($\sim$20 kV) and peak currents ($\sim$100 kA). These conditions can help overcome many of the operational problems that plague other electromagnetic thrusters such as the loss of flow stability and weak levels of plasma compression. For instance, higher operational voltage and current have been shown to enable higher plume compression, acceleration, and even the ability to process and ionize complex molecular propellants. The higher exhaust velocities in particular have even been shown to enable prolonged operation in z-pinch schemes by stabilizing hydromagnetic instabilities through shear-flow effects \cite{underwood2019dynamic,underwood2020schlieren, shumlak2001evidence,shumlak2017increasing}.

The hydromagnetic gun that we have studied is composed of three distinct systems that synergize to produce thrust: (1) an acceleration volume, (2) pulsed-power circuitry, and (3) a neutral gas injector. The acceleration volume of our device consists of a 26 cm long and 5 cm diameter coaxial segment with a 0.5 cm diameter copper cathode and rodded stainless-steel anodes \cite{underwood2017plasma}. Energy is supplied to the gun using a 56 $\mu$F capacitor bank that can be charged up to 20 kV and supply peak currents of up to $\sim$100 kA over $\sim$20 $\mu$s. This current flow induces magnetic field strengths of $\sim$0.1-1 T within the coaxial segment that cause Lorentz forces to accelerate and compress the ionized propellant stream.

To initiate a discharge, neutral gas is injected into the accelerator using a fast rise-rate, variable mass-bit puff valve \cite{loebner2015fast}. This volume is maintained at rarefied conditions ($\sim 10^{-7}$ Torr) between discharge sequences by connecting it to a vacuum chamber that is equipped with a set of cryogenic pumps. The valve operates according to the principle of diamagnetic repulsion and allows the injection of $\sim$10-100 mg of propellant (e.g., N$_{2}$) over $\sim$300 $\mu$s. As the propellant reaches a critical breakdown density, it is rapidly ionized and accelerated out of the thruster.

\subsection{Theory}

Operation of plasma guns has uncovered the existence of two distinct flow patterns that depend on their initial gas loading (Fig.~\ref{fig:1}) \cite{loebner2015evidence, Subramaniam_2018}. When the coaxial volume is pre-filled with gas, a thin current sheet is observed to convect along the electrodes in a process called a ``snow plow’’. This process relies on the propagating conduction zone to collide with upstream mass until the local temperature and electrical conductivity has increased enough to form a shock front. This is the traditional operational mode that is observed in PPTs and allows highly localized current densities to form that ``plow’’ propellant but also rapidly degrade electrodes. If however the thruster is initialized in rarified conditions, as is possible with the Stanford hydromagnetic thruster, a second mode of operation can be observed where a collimated and quasi-steady jet forms \cite{loebner2014high, underwood2019dynamic}.

The properties of the operational modes can be derived by considering a magnetic extension to the Rankine-Hugoniot (RH) theory \cite{cheng1970plasma}. In this model, two distinct current driven ionization waves are predicted if external electrical energy can be added to the flow. A theoretical depiction of the structure and properties of these waves is shown in Fig.~\ref{fig:1}. The magneto-detonation wave, or upper-branch solution, propagates along the length of the accelerator into the unprocessed gas while the magneto-deflagration wave, or lower-branch solution, is quasi-steady and propagates upstream toward the point where propellant is fed in. The validity of the RH model has been investigated in separate studies by measuring the propagation and density characteristics of each wave and comparing that to theoretical predictions. The RH theory was found to agree to within 8\% of measured propagation speeds and correctly predict the deflagration mode is able to achieve higher exhaust velocities while the detonation mode is capable of processing larger instantaneous mass bits \cite{loebner2015evidence}.

Although effective at describing the existence of dual hydromagnetic modes, RH theory does not capture the nuances of how they form. More importantly, simplified conservative jump relations offer little insight into how the distinct modes might smoothly transition into each other. This is an important consideration in the design of an agile propulsion system that may rely on exploiting the properties of these two modes. In the subsequent sections, we present a combination of experiments and simulations to systematically explore the role that gas dynamics plays in selecting the operational mode of the thruster.

\section{Characterization}

\begin{figure*}[tp]
\begin{center}
\includegraphics[width=1\textwidth]{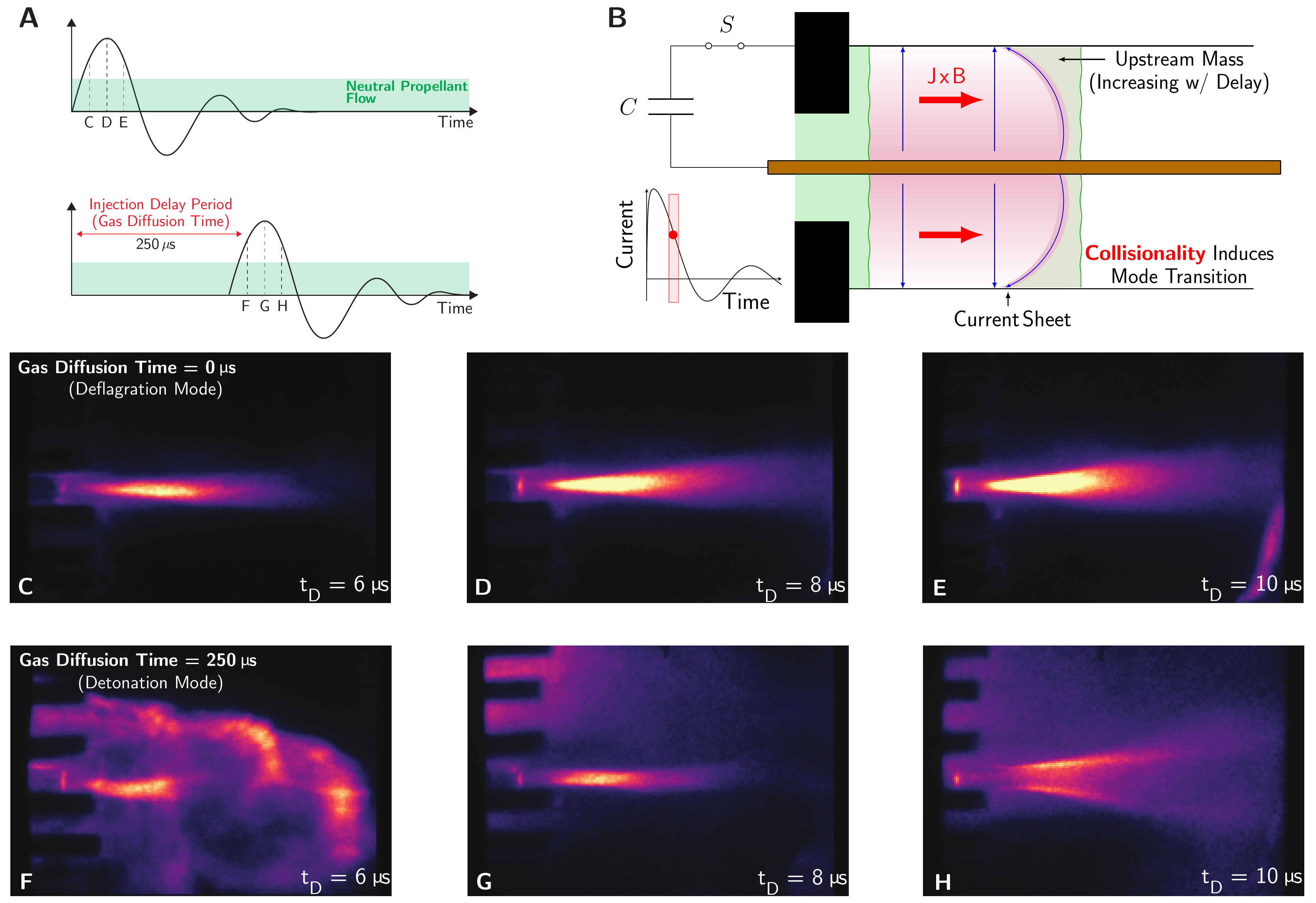}
\end{center}
\caption{The operational characteristics of the thruster can be changed by varying the distribution of propellant within the acceleration volume. By increasing the delay between gas injection (H$_{2}$) and discharge initiation (A), the thruster can smoothly transition (B) from a deflagration wave (C-E) to a detonation wave (F-H). Eventually, increasing upstream mass will induce a collisional shock as the ionization wave propagates through the propellant stream.}\label{fig:2}
\end{figure*}

\subsection{Experimental}

Mode selectivity in the thruster was studied by incorporating additional circuitry to precisely control the initial distribution of propellant before a discharge starts. A combination of H$_{2}$ and N$_{2}$ gas were used to illustrate how fill rates of propellant influence the selectivity and transition between operational modes. A triggerable spark gap was employed to isolate the charged capacitor bank from the thruster until a high voltage trigger pulse was supplied to it. An illustration depicting the timing diagram used to initialize a discharge is shown in Fig.~\ref{fig:2}A. The neutral gas distribution was adjusted within the acceleration volume by changing the time delay between the start of propellant flow and the discharge (referred to as the injection delay or the gas diffusion time). This delay can cause a smooth transition from a deflagration wave to a detonation wave, the latter of which is illustrated in Fig.~\ref{fig:2}B.

The structural evolution of the plasma flow is shown in Fig.~\ref{fig:2}(C-F) for H$_{2}$ gas diffusion times of 0 and 250 $\mu$s respectively with a charging energy of 1.4 kJ. The propellant was puffed into the accelerator with a plenum pressure of 45 psi. Each image is presented in false color and was acquired using an Imacon Ultra8 ICCD camera with a 20 ns exposure time. The discharge times, $t_{D}$, of C-E and F-H were selected to be at the same points during the first half period of the current waveform, as shown in Fig.~\ref{fig:2}A. With no imposed time delay, the operation recovers the behavior of the accelerator operating in puff mode where the electrodes are energized prior to gas being introduced into the thruster, and a deflagration mode is formed. When the H$_{2}$ gas is allowed to expand into the accelerator volume for 250 $\mu$s, the structure of the flow changes. For H$_{2}$, this timescale is sufficiently long to change the initial 10$^{-7}$ vacuum upstream condition throughout the 23 cm long electrode volume. The structure of the propagating front in Fig.~\ref{fig:2}F resembles that of a current sheet. The sensitivity of the ICCD also resolves an expansion wave immediately following the thin shock front. This observation is consistent with the Zeldovich model of a detonation wave as having a thin shock followed by a reaction zone and a trailing expansion wave \cite{von1903theory}. Interestingly, as this detonation wave clears out the channel, a second deflagration occurs during the second swing of the voltage ring-down (Fig.~\ref{fig:2}G and H).

\subsubsection{Stagnation Energy}

\begin{figure*}[tp]
\begin{center}
\includegraphics[width=1\textwidth]{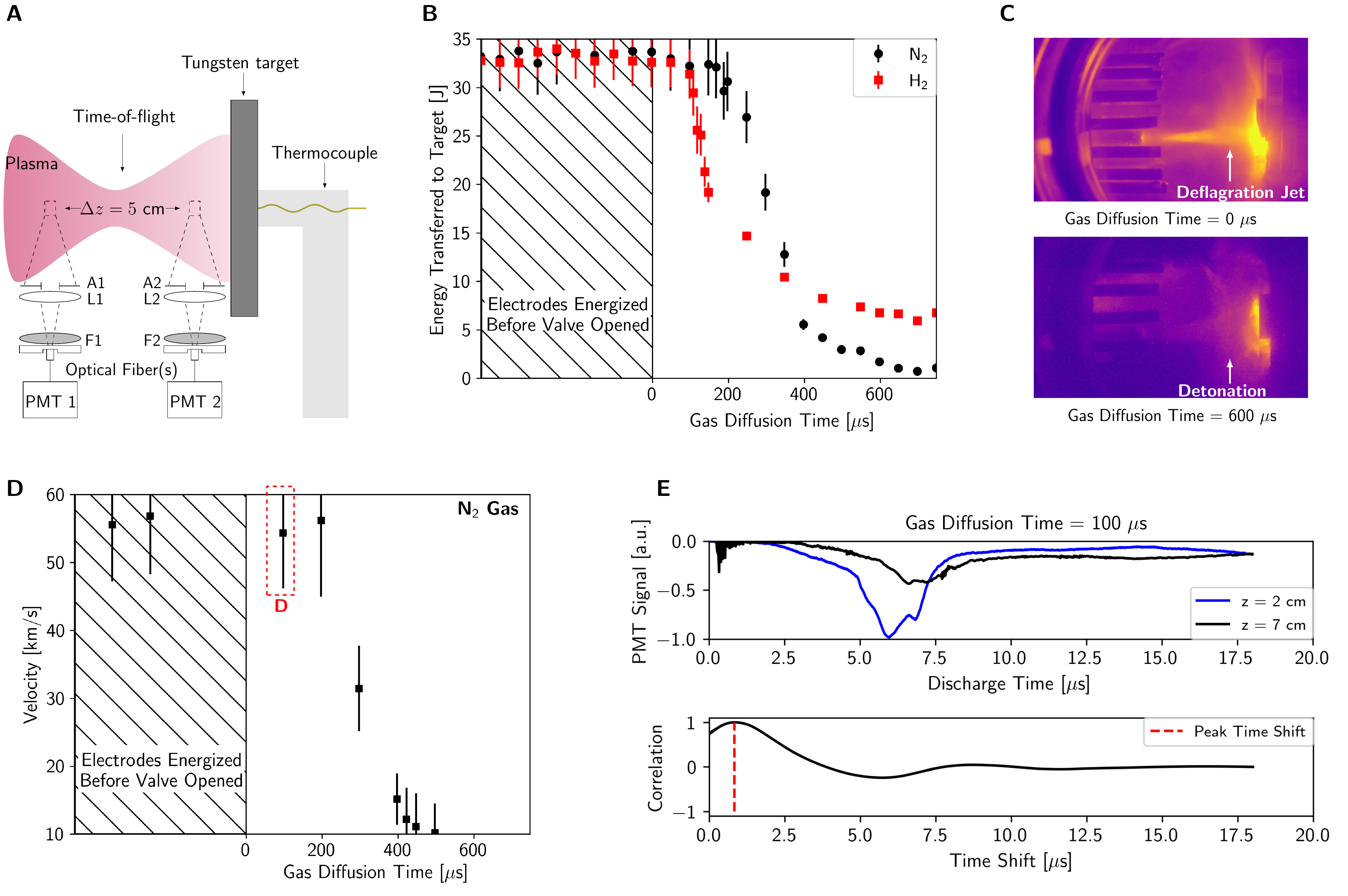}
\end{center}
\caption{(A) Experimental setup used to measure the stagnation energy and exhaust velocity of the plasma plume. (B-C) The energy transfer during stagnation smoothly decays with increasing injection delay. (D-E) Time-of-flight exhaust velocity measurements of the plasma plume. }
\label{fig:3}
\end{figure*}

The directed energy content of the plasma plume is one important metric that can lend insight into the operational transition of the thruster. To assess this energy, a tungsten target was placed 4 cm downstream of the accelerator. The target was 2 cm in diameter, 0.5 cm thick, and mated to a thermally insulating base to center it in the vacuum tube. A 0.5 mm diameter fast type K thermocouple was attached to the backside of the target and connected to an alloy matched vacuum feedthrough to measure the bulk temperature rise, as illustrated in Fig. ~\ref{fig:3}A. The resulting ablative energy measurements as a function of gas diffusion time is shown in Fig.~\ref{fig:3}B. The experiment was charged to 1.4 kJ for each shot and the current and voltage traces remained unchanged with increasing injection delay. The energy transfer was calculated by taking the peak temperature rise experienced by the target of known mass, $E = m c_{p} \Delta T$. Separate trials were conducted for both H$_{2}$ and N$_{2}$ gas with injection delays of up to 800 $\mu$s with an interval of 50 $\mu$s. 

The first set of data points, shown as occurring at negative gas diffusion times (or injection delays) in Fig.~\ref{fig:3}, corresponds to conditions where the electrodes were energized before the gas was injected. With increasing injection delay beyond that, a noticeable reduction in the directed energy content of the flow was recorded. The rate of the energy decay was found to also depend on the type of propellant used. H$_{2}$ was observed to begin decaying sooner and with a larger characteristic slope than N$_{2}$. This suggests that gas dynamic expansion of the propellant plays a prominent role in determining the mode of the thruster. Evidence of a mode transition was also found by taking ICCD images at two extremes of gas diffusion time (0 and 600 $\mu$s), as shown in Fig.~\ref{fig:3}C. 

\subsubsection{Exhaust Velocity}

Time-of-flight measurements were performed to quantify the effect of propellant loading on the acceleration efficiency of the thruster. Measurements with N$_{2}$ as a propellant were made to simulate conditions of interest for air-breathing applications. Two Hammatsu 1P28 photomultiplier tubes (PMTs) were placed 3 cm downstream of the electrodes and 5 cm apart to the visualize the propagation of the plume (Fig.~\ref{fig:3}A). Each PMT was configured to feature a gain of 10$^{6}$, rise times of $\sim$30 ns, and a circular aperture of $\sim$1 mm to limit spatial smearing. The acquired PMT traces for a gas diffusion time of 100 $\mu$s and a charging energy of 1.4 kJ are shown in Fig.~\ref{fig:3}E. From these results, a cross-correlation function was employed to quantify the time delay between detector signals and remove any systematic bias in interpretation (Fig.~\ref{fig:3}E). The measured time delay, $\Delta t$, was used to calculate the plume velocity knowing the spacing between detectors, $V = \text{5 cm}/\Delta t$. 

The plasma flow velocity as a function of injection delay is shown in Fig.~\ref{fig:3}D. As with the temperature rise measurements, the thruster’s exhaust velocity was found to decrease with a similar slope. With a vacuum initial condition, the velocity was found to be approximately 55 km/s (I$_{\text{sp}} \sim 5500$ s). As gas expanded into the accelerator volume, that reduced to 11 km/s (I$_{\text{sp}} \sim 1100$ s). Thus, the change in flow structure and localization is also accompanied by a reduction in the acceleration efficiency. These measurements indicate that propellant injection delay offers a controllable way to smoothly transition from processing larger mass bits (generating more thrust) to more efficiently using propellant (generating more I$_{sp}$).

\begin{figure*}[tp]
\begin{center}
\includegraphics[width=0.98\textwidth]{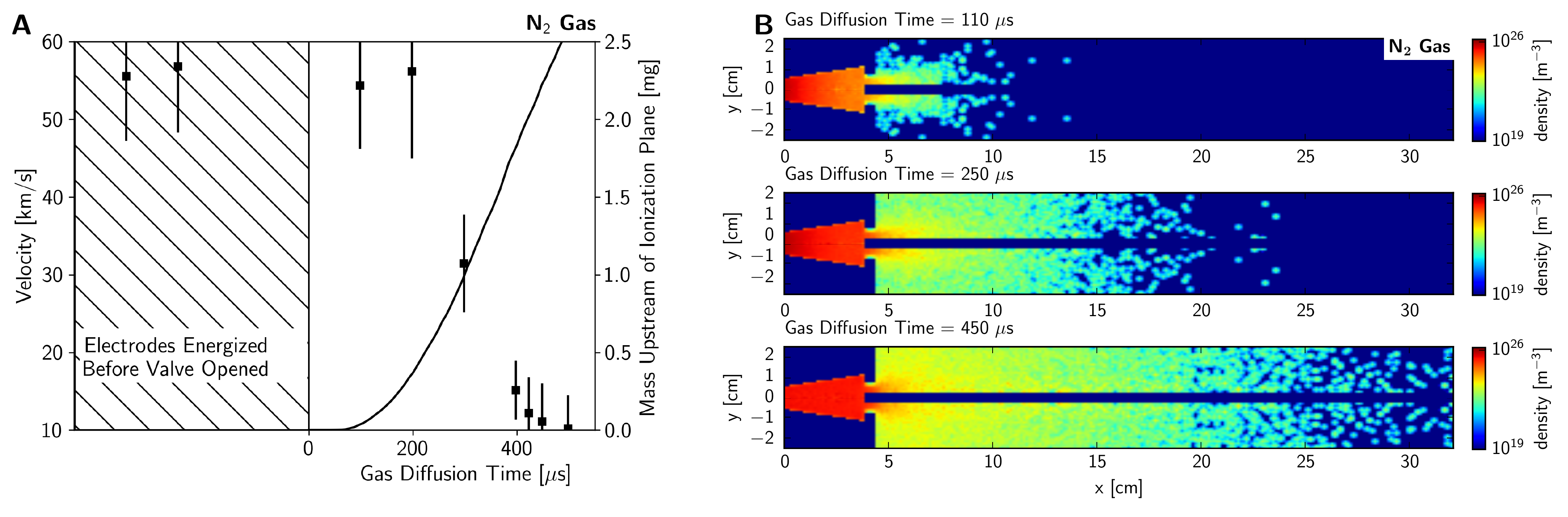}
\end{center}
\caption{(A) Comparison of the measured exhaust velocity and (B) simulated N$_{2}$ propellant distribution in the thruster as a function of the delay between injection and beginning of the discharge. An increasing gas diffusion time (injection delay) results in a more collisional upstream condition for the ionization wave to propagate into.}
\label{fig:4}
\end{figure*}

\subsection{Gas Dynamic Simulations}

A direct simulation Monte Carlo (DSMC) technique was used to explore how gas diffusion time (injection delay) influences the distribution of propellant within the thruster. DSMC codes are popular methods for simulating rarefied flows and have been described in detail elsewhere \cite{boyd2017nonequilibrium, alexander1997direct}. In the system presented here, N$_{2}$ was simulated in 2-D space within an axisymmetric region modeled after the coaxial thruster. Rotational and vibrational degrees of freedom of the gas were ignored. Gas-surface interactions and reflections off of the thruster walls were computed using the Maxwell model with an accommodation coefficient of 0.7. This value was selected based on experimental measurements \cite{eggleton1952thermal}. The gas puff valve was treated as a source reservoir of particles maintained at constant pressure and connected to the nozzle entrance. At each time step, a set of particles was initialized within the reservoir volume, with positions and velocities corresponding to a uniform Maxwellian distribution at a pressure of 45 psi and temperature of 293 K.

To capture the collisional dynamics of gases in DSMC methods, the time step and cell size must be chosen to resolve the minimum collision frequency and collisional mean free path. These requirements become prohibitive in this case as the neutral gas expands from a valve pressure of 45 psi to vacuum ($\sim 10^{-7}$ Torr) conditions. To overcome this, free molecular flow was assumed everywhere, and particles were only allowed to interact with boundary surfaces. For a diatomic molecule like N$_{2}$, the particle flux in the hydrodynamic regime (with collisions) is larger than that of the free molecular regime by a factor of 1.414 \cite{liepmann_1961}.  We therefore expect our DSMC results to underpredict the mass flux into the gun by a similar amount. A complete overview of the numerical procedure and description of the assumptions is provided in Ref. 31.

Results of the neutral gas simulations at select times during the gas expansion process are shown in Fig.~\ref{fig:4}. The simulation region is initiated to be in high vacuum. At t = 0, the valve is opened, and particles begin to enter the domain through the nozzle. The valve remains open for 300 $\mu$s to be consistent with prior experimental characterizations of the puff valve. At each time step, the neutral gas distribution within the gun volume was integrated and recorded to capture its behavior as a function of gas diffusion time. Finally, the integrated mass upstream of the ionization plane is plotted against experimental results of velocity in Fig.~\ref{fig:4}A. Here the ionization plane was assumed to be the first axial point within the acceleration volume where the gas is simultaneously exposed to both electrodes with a pressure that can support breakdown.

\begin{figure*}[tp]
\begin{center}
\includegraphics[width=1\textwidth]{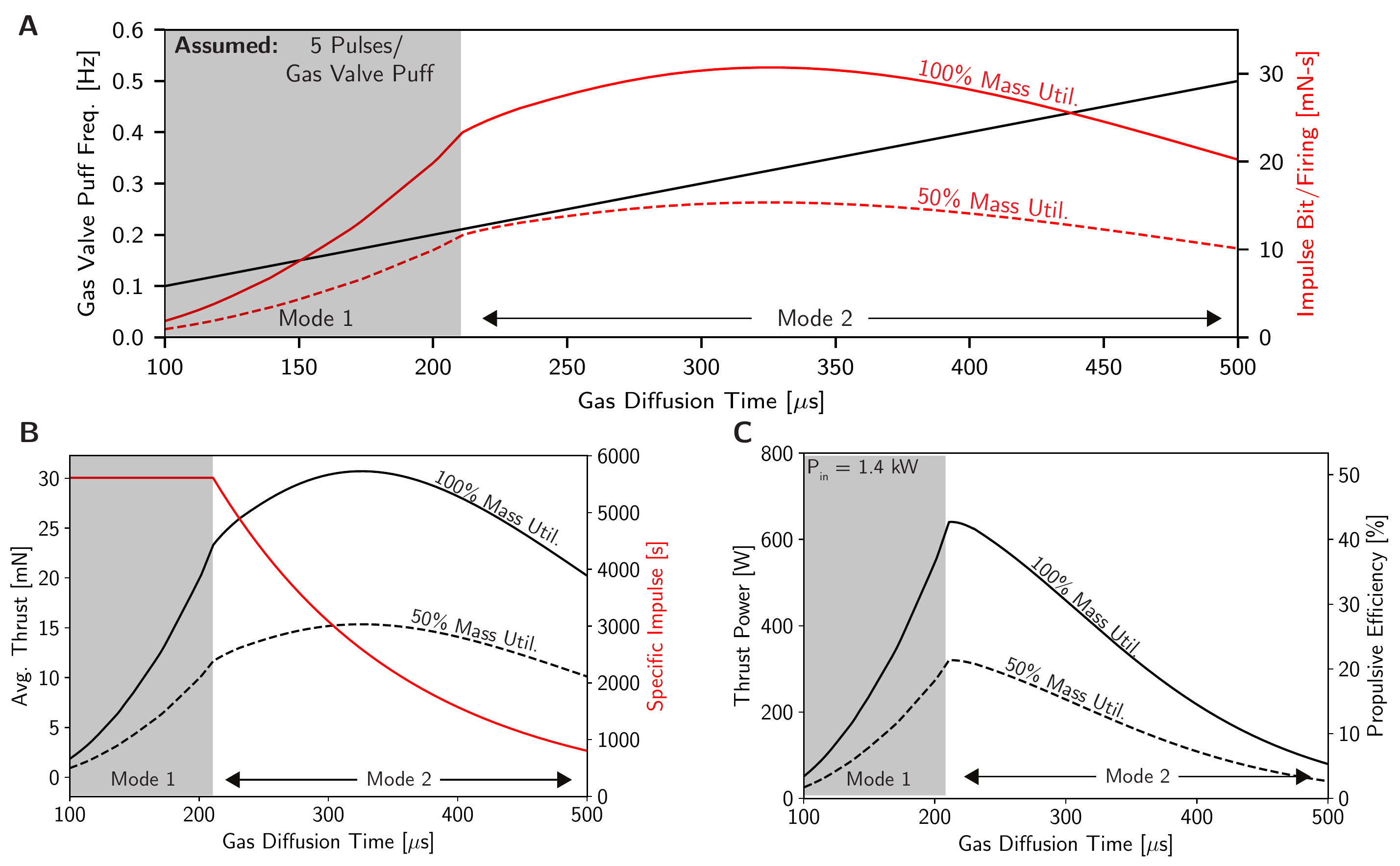}
\end{center}
\caption{Operational characteristics of a proposed multi-mode thruster with an average input power of 1.4 kW. Each gas puff is assumed to supply N$_{2}$ propellant for 5 different $\sim$20 $\mu$s discharge events (lasting $\sim$ 1 ms) with the injected mass optimized for the desired mode of operation. The results assume the measured velocities and processed mass remain constant between discharge events (Fig.~\ref{fig:4}).}
\label{fig:5}
\end{figure*}

The gas dynamic simulations show agreement between the timescales over which gaseous diffusion impacts the experimental measurements (Fig.~\ref{fig:3}) and the slope by which those changes occur. Not only that, they also offer insight into how the operational mode of the thruster can be selected. For instance, a steady magneto-deflagration requires a smooth expansion process to transition from a region of higher pressure and density to a lower one. This means that the flow must remain weakly collisional as it accelerates out of the thruster. When there is no injection delay, the upstream condition is constrained to be 10$^{-6}$-10$^{-7}$ Torr by cryopumps (n$_{\text{n}} \sim 10^{16}$ m$^{-3}$). As the plasma forms at the entrance to the thruster, some fraction of this background gas will be ionized, n$_{\text{p}} \leq \text{n}_{\text{n}}$. Assuming an average convective velocity of $V \sim 10$ km/s and $T \sim 1-10$ eV, the ion-ion collision mean free path ($\lambda_{ii}$) exceeds 1 mm at n$_{\text{p}} \sim 10^{20}$ m$^{-3}$, well above the background propellant levels experienced during plasma acceleration \cite{collidingJets,underwood2019hydromagnetic}. In this way, the neutral gas simulations start predicting collisionally induced mode transitions to become possible with as little as 200 $\mu$s of injection delay while operating with N$_{2}$ propellant. Any further delays past that will impact the strength of the detonation wave \cite{loebner2015evidence}.

\section{Thruster Performance}

The characterization results can be combined to estimate the performance metrics of each operating mode and the tunability range between them. The DSMC results, Fig.~\ref{fig:4}, show that the accelerated mass bit can be tuned depending on the time that propellant is allowed to diffuse into the acceleration volume. This introduces a tradeoff between I$_{\text{sp}}$ and thrust that can be adjusted between discharge pulses of the thruster. For instance, after a gas diffusion time of $\sim$ 100 $\mu$s, the available mass bit of N$_{2}$ (that can be processed) is $\sim$ 0.1 mg while the exhaust velocity is at its highest, $\sim$55 km/s. After a gas diffusion time of $\sim$ 450 $\mu$s however, the exhaust velocity drops to $\sim$ 11 km/s while the available mass bit increases to $>$ 2 mg. 

To characterize its achievable performance, we assume the hydromagnetic thruster operates in a burst mode with an average input power of 1.4 kW. Each thrust event (referred to as a pulse) follows from the characterization results presented in Figs.~\ref{fig:2}-\ref{fig:4}, namely each consumes 1.4 kJ of energy and lasts $\sim$ 20 $\mu$s. Propellant is introduced using the same gas puff valve over $\sim$ 1 ms, however each puff feeds propellant for 5 consecutive pulse events (referred to as a burst). These pulses are then uniformly distributed across the gas puff opening to utilize the propellant. The instantaneous mass bit ($\Delta$m) available for each pulse is assumed to follow from characterization of single pulses in Fig.~\ref{fig:4} and is a function of the gas diffusion time into the accelerator. 


The performance of a hydromagnetic thruster operating on N$_{2}$ propellant with a burst rate of five pulses per valve opening (N$_{p}$ = 5) is shown in Fig.~\ref{fig:5}. The gas puff valve opening frequency (f$_{v}$) is calculated to constrain the average input power of the thruster to be 1.4 kW. From Fig. 5, the thrust, I$_{\text{sp}}$, and propulsive P$_{\text{in}}$/P$_{\text{t}}$ where P$_{\text{t}}$ = 1/2$\Delta$mV$^{2}$f$_{v}$N$_{\text{p}}$) are all tunable between two distinct modes. This illustrates the flexibility that gas dynamics offers to add agility to pulsed plasma thrusters. The characterization of the two modes is extracted and detailed in Table~\ref{tab:1}. The first mode, which occurs at low gas diffusion times ($<$ $\sim$200 $\mu$s), is a deflagration that offers high specific impulse ($\sim$ 5600 s) and propulsion efficiency (up to 43\%). The second mode, a detonation, can be generated by simply allowing the gas to diffuse into the accelerator longer ($\sim$ 350 $\mu$s) and offers high levels of average thrust (up to 31 mN). It is noteworthy that the diameter of the discharge in the detonation mode is $\sim$ 5 cm. In this mode, the thrust density is more than an order of magnitude greater than state-of-the art HETs.

The performance of the thruster also depends on the efficiency that mass is utilized and specifically, the fraction that is accelerated to the measured exhaust velocity ($\eta_{m}$). Performance metrics in Table~\ref{tab:1} represent a theoretical maximum where all of the mass is processed and accelerated to the same velocity ($\eta_{m}$ = 1). The stagnation measurements in Fig.~\ref{fig:3} offer a lower limit on the fraction of mass that can be accelerated. We measured a stagnation energy of $\sim$ 33 J for N$_{2}$ propellant with a gas diffusion time of $\sim$ 200 $\mu$s. Comparing to that calculated thrust power (P$_{t}$), the deflagration mode is capable of operating with a process efficiency of \textit{at least} $\sim$ 10\%. It is important to note that flow dissipation effects, boundary layer effects, and the finite size of the stagnating body (2 cm) mean that all of the energy in the plume cannot be captured (as demonstrated particularly with the larger size of the detonation mode). However, as shown by the difference in flow structure in Fig.~\ref{fig:2}, we expect the capacity of each mode to accelerate propellant to be different. The propagating current sheet in the detonation mode in particular has been shown to act as a plow to accelerate larger instantaneous mass bits to the same exhaust velocity \cite{ziemer2000review}. This difference in flow structure between modes (jet vs. plow) will lead to a further increase in the tunability range between modes and is something that can be instituted on demand by controlling the gas dynamics of the propellant feed.

\begin{table}[tp!]
\setlength{\tabcolsep}{8pt}
    \centering
    \caption{Estimated performance of a hydromagnetic thruster based on the measured velocities and processed mass for a single discharge. The assumed operational parameters include a pulse energy of 1.4 kJ, N$_{2}$ propellant, 100\% mass utilization, and an average input energy of 1.4 kW. Gas diffusion times were selected for each mode by optimizing I$_{\text{sp}}$ and thrust in Fig.~\ref{fig:5}.}
    \vspace{12 pt}
    
    \begin{tabular}{l|c|c}
    \hline
    \textbf{Performance Metric} & \textbf{Deflagration} & \textbf{Detonation} \\
    \hline
        Valve Frequency [Hz] & 0.21 & 0.33\\
        Pulses/Valve Opening & 5 & 5\\
        Pulse Energy [kJ] & 1.4 & 1.4\\
        Mass Bit/Pulse [mg] & 0.42 & 1.2\\
        Average Thrust [mN] & 23  & 31\\
        Isp [s] & 5600 & 2600\\
        Efficiency [\%] & 43 & 26\\
   \hline
    \end{tabular}
    \label{tab:1}
\end{table}

\section{Conclusion}

This manuscript describes a novel hydromagnetic thruster that can operate in two distinct modes. These modes have distinct performance characteristics that can enable the thruster to rapidly transition from optimizing thrust generation to propellant utilization. A theory is presented to describe the structure and evolution of each mode along with motivating the exploration of gas dynamics as a controllable means to enable reconfigurability in thrusters. A combination of rarefied simulations, experiments, and collisionality estimates support this theory and establish baseline performance metrics for a hydromagnetic thruster and a parameter space over which mode transitions can be expected to occur. The ability to dynamically tune both the specific impulse and thrust has exciting implications for the future of space propulsion. This manuscript characterizes a preliminary design of the thruster. Future steps include directly measuring thrust for each operational mode and operating it in a burst configuration.


\section*{Author Contributions}
M.C. and T.U. conceived the idea of a dual mode thruster. T.U. and W.R. performed the experiments. T.U. analyzed the experimental results. W.R. performed the gas dynamic simulations. T.U. and M.C. wrote the manuscript. W.R. described the gas dynamic results. 

\section*{Acknowledgments}
This research is supported by the U.S. Air Force Office of Scientific Research, with Dr. Mitat Birkan as Program Manager. The authors wish to acknowledge Hadland Imaging for generously providing an Imacon ICCD camera to visualize the dynamics of the coaxial plasma flow.

\section*{Data Availability Statement}
The datasets analyzed for this study shall be available upon request.

\bibliographystyle{apsrev4-1} 
\bibliography{test}

\begin{thebibliography}{32}%
\makeatletter
\providecommand \@ifxundefined [1]{%
 \@ifx{#1\undefined}
}%
\providecommand \@ifnum [1]{%
 \ifnum #1\expandafter \@firstoftwo
 \else \expandafter \@secondoftwo
 \fi
}%
\providecommand \@ifx [1]{%
 \ifx #1\expandafter \@firstoftwo
 \else \expandafter \@secondoftwo
 \fi
}%
\providecommand \natexlab [1]{#1}%
\providecommand \enquote  [1]{``#1''}%
\providecommand \bibnamefont  [1]{#1}%
\providecommand \bibfnamefont [1]{#1}%
\providecommand \citenamefont [1]{#1}%
\providecommand \href@noop [0]{\@secondoftwo}%
\providecommand \href [0]{\begingroup \@sanitize@url \@href}%
\providecommand \@href[1]{\@@startlink{#1}\@@href}%
\providecommand \@@href[1]{\endgroup#1\@@endlink}%
\providecommand \@sanitize@url [0]{\catcode `\\12\catcode `\$12\catcode
  `\&12\catcode `\#12\catcode `\^12\catcode `\_12\catcode `\%12\relax}%
\providecommand \@@startlink[1]{}%
\providecommand \@@endlink[0]{}%
\providecommand \url  [0]{\begingroup\@sanitize@url \@url }%
\providecommand \@url [1]{\endgroup\@href {#1}{\urlprefix }}%
\providecommand \urlprefix  [0]{URL }%
\providecommand \Eprint [0]{\href }%
\providecommand \doibase [0]{http://dx.doi.org/}%
\providecommand \selectlanguage [0]{\@gobble}%
\providecommand \bibinfo  [0]{\@secondoftwo}%
\providecommand \bibfield  [0]{\@secondoftwo}%
\providecommand \translation [1]{[#1]}%
\providecommand \BibitemOpen [0]{}%
\providecommand \bibitemStop [0]{}%
\providecommand \bibitemNoStop [0]{.\EOS\space}%
\providecommand \EOS [0]{\spacefactor3000\relax}%
\providecommand \BibitemShut  [1]{\csname bibitem#1\endcsname}%
\let\auto@bib@innerbib\@empty
\bibitem [{\citenamefont {Peukert}\ and\ \citenamefont
  {Wollenhaupt}(2014)}]{peukert2014ohb}%
  \BibitemOpen
  \bibfield  {author} {\bibinfo {author} {\bibfnamefont {M.}~\bibnamefont
  {Peukert}}\ and\ \bibinfo {author} {\bibfnamefont {B.}~\bibnamefont
  {Wollenhaupt}},\ }in\ \href@noop {} {\emph {\bibinfo {booktitle} {EPIC
  Workshop, Brussels}}}\ (\bibinfo {year} {2014})\BibitemShut {NoStop}%
\bibitem [{\citenamefont {Mazouffre}(2016)}]{mazouffre2016electric}%
  \BibitemOpen
  \bibfield  {author} {\bibinfo {author} {\bibfnamefont {S.}~\bibnamefont
  {Mazouffre}},\ }\href@noop {} {\bibfield  {journal} {\bibinfo  {journal}
  {Plasma Sources Science and Technology}\ }\textbf {\bibinfo {volume} {25}},\
  \bibinfo {pages} {033002} (\bibinfo {year} {2016})}\BibitemShut {NoStop}%
\bibitem [{\citenamefont {Cifali}\ \emph {et~al.}(2011)\citenamefont {Cifali},
  \citenamefont {Misuri}, \citenamefont {Rossetti}, \citenamefont {Andrenucci},
  \citenamefont {Valentian},\ and\ \citenamefont
  {Feili}}]{cifali2011preliminary}%
  \BibitemOpen
  \bibfield  {author} {\bibinfo {author} {\bibfnamefont {G.}~\bibnamefont
  {Cifali}}, \bibinfo {author} {\bibfnamefont {T.}~\bibnamefont {Misuri}},
  \bibinfo {author} {\bibfnamefont {P.}~\bibnamefont {Rossetti}}, \bibinfo
  {author} {\bibfnamefont {M.}~\bibnamefont {Andrenucci}}, \bibinfo {author}
  {\bibfnamefont {D.}~\bibnamefont {Valentian}}, \ and\ \bibinfo {author}
  {\bibfnamefont {D.}~\bibnamefont {Feili}},\ }in\ \href@noop {} {\emph
  {\bibinfo {booktitle} {47th AIAA/ASME/SAE/ASEE Joint Propulsion Conference \&
  Exhibit}}}\ (\bibinfo {year} {2011})\ p.\ \bibinfo {pages} {6073}\BibitemShut
  {NoStop}%
\bibitem [{\citenamefont {Gurciullo}\ \emph {et~al.}(2019)\citenamefont
  {Gurciullo}, \citenamefont {Fabris},\ and\ \citenamefont
  {Cappelli}}]{gurciullo2019ion}%
  \BibitemOpen
  \bibfield  {author} {\bibinfo {author} {\bibfnamefont {A.}~\bibnamefont
  {Gurciullo}}, \bibinfo {author} {\bibfnamefont {A.~L.}\ \bibnamefont
  {Fabris}}, \ and\ \bibinfo {author} {\bibfnamefont {M.~A.}\ \bibnamefont
  {Cappelli}},\ }\href@noop {} {\bibfield  {journal} {\bibinfo  {journal}
  {Journal of Physics D: Applied Physics}\ }\textbf {\bibinfo {volume} {52}},\
  \bibinfo {pages} {464003} (\bibinfo {year} {2019})}\BibitemShut {NoStop}%
\bibitem [{\citenamefont {Dietz}\ \emph {et~al.}(2019)\citenamefont {Dietz},
  \citenamefont {G{\"a}rtner}, \citenamefont {Koch}, \citenamefont
  {K{\"o}hler}, \citenamefont {Teng}, \citenamefont {Schreiner}, \citenamefont
  {Holste},\ and\ \citenamefont {Klar}}]{dietz2019molecular}%
  \BibitemOpen
  \bibfield  {author} {\bibinfo {author} {\bibfnamefont {P.}~\bibnamefont
  {Dietz}}, \bibinfo {author} {\bibfnamefont {W.}~\bibnamefont {G{\"a}rtner}},
  \bibinfo {author} {\bibfnamefont {Q.}~\bibnamefont {Koch}}, \bibinfo {author}
  {\bibfnamefont {P.~E.}\ \bibnamefont {K{\"o}hler}}, \bibinfo {author}
  {\bibfnamefont {Y.}~\bibnamefont {Teng}}, \bibinfo {author} {\bibfnamefont
  {P.~R.}\ \bibnamefont {Schreiner}}, \bibinfo {author} {\bibfnamefont
  {K.}~\bibnamefont {Holste}}, \ and\ \bibinfo {author} {\bibfnamefont {P.~J.}\
  \bibnamefont {Klar}},\ }\href@noop {} {\bibfield  {journal} {\bibinfo
  {journal} {Plasma Sources Science and Technology}\ }\textbf {\bibinfo
  {volume} {28}},\ \bibinfo {pages} {084001} (\bibinfo {year}
  {2019})}\BibitemShut {NoStop}%
\bibitem [{\citenamefont {Romano}\ \emph {et~al.}(2018)\citenamefont {Romano},
  \citenamefont {Massuti-Ballester}, \citenamefont {Binder}, \citenamefont
  {Herdrich}, \citenamefont {Fasoulas},\ and\ \citenamefont
  {Sch{\"o}nherr}}]{romano2018system}%
  \BibitemOpen
  \bibfield  {author} {\bibinfo {author} {\bibfnamefont {F.}~\bibnamefont
  {Romano}}, \bibinfo {author} {\bibfnamefont {B.}~\bibnamefont
  {Massuti-Ballester}}, \bibinfo {author} {\bibfnamefont {T.}~\bibnamefont
  {Binder}}, \bibinfo {author} {\bibfnamefont {G.}~\bibnamefont {Herdrich}},
  \bibinfo {author} {\bibfnamefont {S.}~\bibnamefont {Fasoulas}}, \ and\
  \bibinfo {author} {\bibfnamefont {T.}~\bibnamefont {Sch{\"o}nherr}},\
  }\href@noop {} {\bibfield  {journal} {\bibinfo  {journal} {Acta
  Astronautica}\ }\textbf {\bibinfo {volume} {147}},\ \bibinfo {pages} {114}
  (\bibinfo {year} {2018})}\BibitemShut {NoStop}%
\bibitem [{\citenamefont {Romano}\ \emph {et~al.}(2020)\citenamefont {Romano},
  \citenamefont {Chan}, \citenamefont {Herdrich}, \citenamefont {Traub},
  \citenamefont {Fasoulas}, \citenamefont {Roberts}, \citenamefont {Smith},
  \citenamefont {Edmondson}, \citenamefont {Haigh}, \citenamefont {Crisp} \emph
  {et~al.}}]{romano2020rf}%
  \BibitemOpen
  \bibfield  {author} {\bibinfo {author} {\bibfnamefont {F.}~\bibnamefont
  {Romano}}, \bibinfo {author} {\bibfnamefont {Y.-A.}\ \bibnamefont {Chan}},
  \bibinfo {author} {\bibfnamefont {G.}~\bibnamefont {Herdrich}}, \bibinfo
  {author} {\bibfnamefont {C.}~\bibnamefont {Traub}}, \bibinfo {author}
  {\bibfnamefont {S.}~\bibnamefont {Fasoulas}}, \bibinfo {author}
  {\bibfnamefont {P.}~\bibnamefont {Roberts}}, \bibinfo {author} {\bibfnamefont
  {K.}~\bibnamefont {Smith}}, \bibinfo {author} {\bibfnamefont
  {S.}~\bibnamefont {Edmondson}}, \bibinfo {author} {\bibfnamefont
  {S.}~\bibnamefont {Haigh}}, \bibinfo {author} {\bibfnamefont
  {N.}~\bibnamefont {Crisp}},  \emph {et~al.},\ }\href@noop {} {\bibfield
  {journal} {\bibinfo  {journal} {Acta Astronautica}\ }\textbf {\bibinfo
  {volume} {176}},\ \bibinfo {pages} {476} (\bibinfo {year}
  {2020})}\BibitemShut {NoStop}%
\bibitem [{\citenamefont {Ziemer}\ and\ \citenamefont
  {Petr}(2002)}]{ziemer2002performance}%
  \BibitemOpen
  \bibfield  {author} {\bibinfo {author} {\bibfnamefont {J.}~\bibnamefont
  {Ziemer}}\ and\ \bibinfo {author} {\bibfnamefont {R.}~\bibnamefont {Petr}},\
  }in\ \href@noop {} {\emph {\bibinfo {booktitle} {38th AIAA/ASME/SAE/ASEE
  Joint Propulsion Conference \& Exhibit}}}\ (\bibinfo {year} {2002})\ p.\
  \bibinfo {pages} {4273}\BibitemShut {NoStop}%
\bibitem [{\citenamefont {Larson}\ \emph {et~al.}(1965)\citenamefont {Larson},
  \citenamefont {Gooding}, \citenamefont {Hayworth},\ and\ \citenamefont
  {Ashby}}]{larson1965energy}%
  \BibitemOpen
  \bibfield  {author} {\bibinfo {author} {\bibfnamefont {A.}~\bibnamefont
  {Larson}}, \bibinfo {author} {\bibfnamefont {T.}~\bibnamefont {Gooding}},
  \bibinfo {author} {\bibfnamefont {B.}~\bibnamefont {Hayworth}}, \ and\
  \bibinfo {author} {\bibfnamefont {D.}~\bibnamefont {Ashby}},\ }\href@noop {}
  {\bibfield  {journal} {\bibinfo  {journal} {AIAA Journal}\ }\textbf {\bibinfo
  {volume} {3}},\ \bibinfo {pages} {977} (\bibinfo {year} {1965})}\BibitemShut
  {NoStop}%
\bibitem [{\citenamefont {Mather}(1964)}]{mather1964investigation}%
  \BibitemOpen
  \bibfield  {author} {\bibinfo {author} {\bibfnamefont {J.}~\bibnamefont
  {Mather}},\ }\href@noop {} {\bibfield  {journal} {\bibinfo  {journal} {The
  Physics of Fluids}\ }\textbf {\bibinfo {volume} {7}},\ \bibinfo {pages} {S28}
  (\bibinfo {year} {1964})}\BibitemShut {NoStop}%
\bibitem [{\citenamefont {Loebner}\ \emph
  {et~al.}(2015{\natexlab{a}})\citenamefont {Loebner}, \citenamefont
  {Underwood},\ and\ \citenamefont {Cappelli}}]{loebner2015evidence}%
  \BibitemOpen
  \bibfield  {author} {\bibinfo {author} {\bibfnamefont {K.~T.}\ \bibnamefont
  {Loebner}}, \bibinfo {author} {\bibfnamefont {T.~C.}\ \bibnamefont
  {Underwood}}, \ and\ \bibinfo {author} {\bibfnamefont {M.~A.}\ \bibnamefont
  {Cappelli}},\ }\href@noop {} {\bibfield  {journal} {\bibinfo  {journal}
  {Physical review letters}\ }\textbf {\bibinfo {volume} {115}},\ \bibinfo
  {pages} {175001} (\bibinfo {year} {2015}{\natexlab{a}})}\BibitemShut
  {NoStop}%
\bibitem [{\citenamefont {Marshall}(1960)}]{marshall1960performance}%
  \BibitemOpen
  \bibfield  {author} {\bibinfo {author} {\bibfnamefont {J.}~\bibnamefont
  {Marshall}},\ }\href@noop {} {\bibfield  {journal} {\bibinfo  {journal}
  {Physics of Fluids (US)}\ }\textbf {\bibinfo {volume} {3}} (\bibinfo {year}
  {1960})}\BibitemShut {NoStop}%
\bibitem [{\citenamefont {Underwood}\ \emph {et~al.}(2017)\citenamefont
  {Underwood}, \citenamefont {Loebner},\ and\ \citenamefont
  {Cappelli}}]{underwood2017plasma}%
  \BibitemOpen
  \bibfield  {author} {\bibinfo {author} {\bibfnamefont {T.~C.}\ \bibnamefont
  {Underwood}}, \bibinfo {author} {\bibfnamefont {K.~T.}\ \bibnamefont
  {Loebner}}, \ and\ \bibinfo {author} {\bibfnamefont {M.~A.}\ \bibnamefont
  {Cappelli}},\ }\href@noop {} {\bibfield  {journal} {\bibinfo  {journal} {High
  energy density physics}\ }\textbf {\bibinfo {volume} {23}},\ \bibinfo {pages}
  {73} (\bibinfo {year} {2017})}\BibitemShut {NoStop}%
\bibitem [{\citenamefont {Thio}\ \emph {et~al.}(2019)\citenamefont {Thio},
  \citenamefont {Hsu}, \citenamefont {Witherspoon}, \citenamefont {Cruz},
  \citenamefont {Case}, \citenamefont {Langendorf}, \citenamefont {Yates},
  \citenamefont {Dunn}, \citenamefont {Cassibry}, \citenamefont {Samulyak}
  \emph {et~al.}}]{thio2019plasma}%
  \BibitemOpen
  \bibfield  {author} {\bibinfo {author} {\bibfnamefont {Y.~F.}\ \bibnamefont
  {Thio}}, \bibinfo {author} {\bibfnamefont {S.~C.}\ \bibnamefont {Hsu}},
  \bibinfo {author} {\bibfnamefont {F.~D.}\ \bibnamefont {Witherspoon}},
  \bibinfo {author} {\bibfnamefont {E.}~\bibnamefont {Cruz}}, \bibinfo {author}
  {\bibfnamefont {A.}~\bibnamefont {Case}}, \bibinfo {author} {\bibfnamefont
  {S.}~\bibnamefont {Langendorf}}, \bibinfo {author} {\bibfnamefont
  {K.}~\bibnamefont {Yates}}, \bibinfo {author} {\bibfnamefont
  {J.}~\bibnamefont {Dunn}}, \bibinfo {author} {\bibfnamefont {J.}~\bibnamefont
  {Cassibry}}, \bibinfo {author} {\bibfnamefont {R.}~\bibnamefont {Samulyak}},
  \emph {et~al.},\ }\href@noop {} {\bibfield  {journal} {\bibinfo  {journal}
  {Fusion Science and Technology}\ }\textbf {\bibinfo {volume} {75}},\ \bibinfo
  {pages} {581} (\bibinfo {year} {2019})}\BibitemShut {NoStop}%
\bibitem [{\citenamefont {Shumlak}\ \emph {et~al.}(2001)\citenamefont
  {Shumlak}, \citenamefont {Golingo}, \citenamefont {Nelson},\ and\
  \citenamefont {Den~Hartog}}]{shumlak2001evidence}%
  \BibitemOpen
  \bibfield  {author} {\bibinfo {author} {\bibfnamefont {U.}~\bibnamefont
  {Shumlak}}, \bibinfo {author} {\bibfnamefont {R.}~\bibnamefont {Golingo}},
  \bibinfo {author} {\bibfnamefont {B.}~\bibnamefont {Nelson}}, \ and\ \bibinfo
  {author} {\bibfnamefont {D.}~\bibnamefont {Den~Hartog}},\ }\href@noop {}
  {\bibfield  {journal} {\bibinfo  {journal} {Physical Review Letters}\
  }\textbf {\bibinfo {volume} {87}},\ \bibinfo {pages} {205005} (\bibinfo
  {year} {2001})}\BibitemShut {NoStop}%
\bibitem [{\citenamefont {Underwood}\ \emph
  {et~al.}(2019{\natexlab{a}})\citenamefont {Underwood}, \citenamefont
  {Subramaniam}, \citenamefont {Riedel}, \citenamefont {Raja},\ and\
  \citenamefont {Cappelli}}]{UNDERWOOD201997}%
  \BibitemOpen
  \bibfield  {author} {\bibinfo {author} {\bibfnamefont {T.~C.}\ \bibnamefont
  {Underwood}}, \bibinfo {author} {\bibfnamefont {V.}~\bibnamefont
  {Subramaniam}}, \bibinfo {author} {\bibfnamefont {W.~M.}\ \bibnamefont
  {Riedel}}, \bibinfo {author} {\bibfnamefont {L.~L.}\ \bibnamefont {Raja}}, \
  and\ \bibinfo {author} {\bibfnamefont {M.~A.}\ \bibnamefont {Cappelli}},\
  }\href {\doibase https://doi.org/10.1016/j.fusengdes.2019.04.088} {\bibfield
  {journal} {\bibinfo  {journal} {Fusion Engineering and Design}\ }\textbf
  {\bibinfo {volume} {144}},\ \bibinfo {pages} {97 } (\bibinfo {year}
  {2019}{\natexlab{a}})}\BibitemShut {NoStop}%
\bibitem [{\citenamefont {Zhang}\ \emph {et~al.}(2019)\citenamefont {Zhang},
  \citenamefont {Shumlak}, \citenamefont {Nelson}, \citenamefont {Golingo},
  \citenamefont {Weber}, \citenamefont {Stepanov}, \citenamefont {Claveau},
  \citenamefont {Forbes}, \citenamefont {Draper}, \citenamefont {Mitrani} \emph
  {et~al.}}]{zhang2019sustained}%
  \BibitemOpen
  \bibfield  {author} {\bibinfo {author} {\bibfnamefont {Y.}~\bibnamefont
  {Zhang}}, \bibinfo {author} {\bibfnamefont {U.}~\bibnamefont {Shumlak}},
  \bibinfo {author} {\bibfnamefont {B.}~\bibnamefont {Nelson}}, \bibinfo
  {author} {\bibfnamefont {R.}~\bibnamefont {Golingo}}, \bibinfo {author}
  {\bibfnamefont {T.}~\bibnamefont {Weber}}, \bibinfo {author} {\bibfnamefont
  {A.}~\bibnamefont {Stepanov}}, \bibinfo {author} {\bibfnamefont
  {E.}~\bibnamefont {Claveau}}, \bibinfo {author} {\bibfnamefont
  {E.}~\bibnamefont {Forbes}}, \bibinfo {author} {\bibfnamefont
  {Z.}~\bibnamefont {Draper}}, \bibinfo {author} {\bibfnamefont
  {J.}~\bibnamefont {Mitrani}},  \emph {et~al.},\ }\href@noop {} {\bibfield
  {journal} {\bibinfo  {journal} {Physical review letters}\ }\textbf {\bibinfo
  {volume} {122}},\ \bibinfo {pages} {135001} (\bibinfo {year}
  {2019})}\BibitemShut {NoStop}%
\bibitem [{\citenamefont {Ziemer}(2000)}]{ziemer2000review}%
  \BibitemOpen
  \bibfield  {author} {\bibinfo {author} {\bibfnamefont {J.~K.}\ \bibnamefont
  {Ziemer}},\ }\href@noop {} {\bibfield  {journal} {\bibinfo  {journal}
  {Electric Propulsion and Plasma Dynamics Lab report, Princeton University}\ }
  (\bibinfo {year} {2000})}\BibitemShut {NoStop}%
\bibitem [{\citenamefont {Underwood}\ \emph
  {et~al.}(2019{\natexlab{b}})\citenamefont {Underwood}, \citenamefont
  {Loebner}, \citenamefont {Miller},\ and\ \citenamefont
  {Cappelli}}]{underwood2019dynamic}%
  \BibitemOpen
  \bibfield  {author} {\bibinfo {author} {\bibfnamefont {T.~C.}\ \bibnamefont
  {Underwood}}, \bibinfo {author} {\bibfnamefont {K.~T.}\ \bibnamefont
  {Loebner}}, \bibinfo {author} {\bibfnamefont {V.~A.}\ \bibnamefont {Miller}},
  \ and\ \bibinfo {author} {\bibfnamefont {M.~A.}\ \bibnamefont {Cappelli}},\
  }\href@noop {} {\bibfield  {journal} {\bibinfo  {journal} {Scientific
  reports}\ }\textbf {\bibinfo {volume} {9}},\ \bibinfo {pages} {1} (\bibinfo
  {year} {2019}{\natexlab{b}})}\BibitemShut {NoStop}%
\bibitem [{\citenamefont {Underwood}\ \emph {et~al.}(2020)\citenamefont
  {Underwood}, \citenamefont {Loebner}, \citenamefont {Miller},\ and\
  \citenamefont {Cappelli}}]{underwood2020schlieren}%
  \BibitemOpen
  \bibfield  {author} {\bibinfo {author} {\bibfnamefont {T.~C.}\ \bibnamefont
  {Underwood}}, \bibinfo {author} {\bibfnamefont {K.~T.}\ \bibnamefont
  {Loebner}}, \bibinfo {author} {\bibfnamefont {V.~A.}\ \bibnamefont {Miller}},
  \ and\ \bibinfo {author} {\bibfnamefont {M.~A.}\ \bibnamefont {Cappelli}},\
  }\href@noop {} {\bibfield  {journal} {\bibinfo  {journal} {Experiments in
  Fluids}\ }\textbf {\bibinfo {volume} {61}},\ \bibinfo {pages} {17} (\bibinfo
  {year} {2020})}\BibitemShut {NoStop}%
\bibitem [{\citenamefont {Shumlak}\ \emph {et~al.}(2017)\citenamefont
  {Shumlak}, \citenamefont {Nelson}, \citenamefont {Claveau}, \citenamefont
  {Forbes}, \citenamefont {Golingo}, \citenamefont {Hughes}, \citenamefont
  {Oberto}, \citenamefont {Ross},\ and\ \citenamefont
  {Weber}}]{shumlak2017increasing}%
  \BibitemOpen
  \bibfield  {author} {\bibinfo {author} {\bibfnamefont {U.}~\bibnamefont
  {Shumlak}}, \bibinfo {author} {\bibfnamefont {B.}~\bibnamefont {Nelson}},
  \bibinfo {author} {\bibfnamefont {E.}~\bibnamefont {Claveau}}, \bibinfo
  {author} {\bibfnamefont {E.}~\bibnamefont {Forbes}}, \bibinfo {author}
  {\bibfnamefont {R.}~\bibnamefont {Golingo}}, \bibinfo {author} {\bibfnamefont
  {M.}~\bibnamefont {Hughes}}, \bibinfo {author} {\bibfnamefont
  {R.}~\bibnamefont {Oberto}}, \bibinfo {author} {\bibfnamefont
  {M.}~\bibnamefont {Ross}}, \ and\ \bibinfo {author} {\bibfnamefont
  {T.}~\bibnamefont {Weber}},\ }\href@noop {} {\bibfield  {journal} {\bibinfo
  {journal} {Physics of Plasmas}\ }\textbf {\bibinfo {volume} {24}},\ \bibinfo
  {pages} {055702} (\bibinfo {year} {2017})}\BibitemShut {NoStop}%
\bibitem [{\citenamefont {Loebner}\ \emph
  {et~al.}(2015{\natexlab{b}})\citenamefont {Loebner}, \citenamefont
  {Underwood},\ and\ \citenamefont {Cappelli}}]{loebner2015fast}%
  \BibitemOpen
  \bibfield  {author} {\bibinfo {author} {\bibfnamefont {K.~T.}\ \bibnamefont
  {Loebner}}, \bibinfo {author} {\bibfnamefont {T.~C.}\ \bibnamefont
  {Underwood}}, \ and\ \bibinfo {author} {\bibfnamefont {M.~A.}\ \bibnamefont
  {Cappelli}},\ }\href@noop {} {\bibfield  {journal} {\bibinfo  {journal}
  {Review of Scientific Instruments}\ }\textbf {\bibinfo {volume} {86}},\
  \bibinfo {pages} {063503} (\bibinfo {year} {2015}{\natexlab{b}})}\BibitemShut
  {NoStop}%
\bibitem [{\citenamefont {Subramaniam}\ \emph {et~al.}(2018)\citenamefont
  {Subramaniam}, \citenamefont {Underwood}, \citenamefont {Raja},\ and\
  \citenamefont {Cappelli}}]{Subramaniam_2018}%
  \BibitemOpen
  \bibfield  {author} {\bibinfo {author} {\bibfnamefont {V.}~\bibnamefont
  {Subramaniam}}, \bibinfo {author} {\bibfnamefont {T.~C.}\ \bibnamefont
  {Underwood}}, \bibinfo {author} {\bibfnamefont {L.~L.}\ \bibnamefont {Raja}},
  \ and\ \bibinfo {author} {\bibfnamefont {M.~A.}\ \bibnamefont {Cappelli}},\
  }\href {\doibase 10.1088/1361-6595/aaabec} {\bibfield  {journal} {\bibinfo
  {journal} {Plasma Sources Science and Technology}\ }\textbf {\bibinfo
  {volume} {27}},\ \bibinfo {pages} {025016} (\bibinfo {year}
  {2018})}\BibitemShut {NoStop}%
\bibitem [{\citenamefont {Loebner}\ \emph {et~al.}(2014)\citenamefont
  {Loebner}, \citenamefont {Wang}, \citenamefont {Poehlmann}, \citenamefont
  {Watanabe},\ and\ \citenamefont {Cappelli}}]{loebner2014high}%
  \BibitemOpen
  \bibfield  {author} {\bibinfo {author} {\bibfnamefont {K.~T.}\ \bibnamefont
  {Loebner}}, \bibinfo {author} {\bibfnamefont {B.~C.}\ \bibnamefont {Wang}},
  \bibinfo {author} {\bibfnamefont {F.~R.}\ \bibnamefont {Poehlmann}}, \bibinfo
  {author} {\bibfnamefont {Y.}~\bibnamefont {Watanabe}}, \ and\ \bibinfo
  {author} {\bibfnamefont {M.~A.}\ \bibnamefont {Cappelli}},\ }\href@noop {}
  {\bibfield  {journal} {\bibinfo  {journal} {IEEE Transactions on Plasma
  Science}\ }\textbf {\bibinfo {volume} {42}},\ \bibinfo {pages} {2500}
  (\bibinfo {year} {2014})}\BibitemShut {NoStop}%
\bibitem [{\citenamefont {Cheng}(1970)}]{cheng1970plasma}%
  \BibitemOpen
  \bibfield  {author} {\bibinfo {author} {\bibfnamefont {D.~Y.}\ \bibnamefont
  {Cheng}},\ }\href@noop {} {\bibfield  {journal} {\bibinfo  {journal} {Nuclear
  Fusion}\ }\textbf {\bibinfo {volume} {10}},\ \bibinfo {pages} {305} (\bibinfo
  {year} {1970})}\BibitemShut {NoStop}%
\bibitem [{\citenamefont {von Neumann}\ and\ \citenamefont
  {Taub}(1903)}]{von1903theory}%
  \BibitemOpen
  \bibfield  {author} {\bibinfo {author} {\bibfnamefont {J.}~\bibnamefont {von
  Neumann}}\ and\ \bibinfo {author} {\bibfnamefont {A.}~\bibnamefont {Taub}},\
  }\href@noop {} {\bibfield  {journal} {\bibinfo  {journal} {John von Neumann:
  Collected Works}\ }\textbf {\bibinfo {volume} {1957}} (\bibinfo {year}
  {1903})}\BibitemShut {NoStop}%
\bibitem [{\citenamefont {Boyd}\ and\ \citenamefont
  {Schwartzentruber}(2017)}]{boyd2017nonequilibrium}%
  \BibitemOpen
  \bibfield  {author} {\bibinfo {author} {\bibfnamefont {I.~D.}\ \bibnamefont
  {Boyd}}\ and\ \bibinfo {author} {\bibfnamefont {T.~E.}\ \bibnamefont
  {Schwartzentruber}},\ }\href@noop {} {\emph {\bibinfo {title} {Nonequilibrium
  Gas Dynamics and Molecular Simulation}}},\ Vol.~\bibinfo {volume} {42}\
  (\bibinfo  {publisher} {Cambridge University Press},\ \bibinfo {year}
  {2017})\BibitemShut {NoStop}%
\bibitem [{\citenamefont {Alexander}\ and\ \citenamefont
  {Garcia}(1997)}]{alexander1997direct}%
  \BibitemOpen
  \bibfield  {author} {\bibinfo {author} {\bibfnamefont {F.~J.}\ \bibnamefont
  {Alexander}}\ and\ \bibinfo {author} {\bibfnamefont {A.~L.}\ \bibnamefont
  {Garcia}},\ }\href@noop {} {\bibfield  {journal} {\bibinfo  {journal}
  {Computers in Physics}\ }\textbf {\bibinfo {volume} {11}},\ \bibinfo {pages}
  {588} (\bibinfo {year} {1997})}\BibitemShut {NoStop}%
\bibitem [{\citenamefont {Eggleton}\ and\ \citenamefont
  {Tompkins}(1952)}]{eggleton1952thermal}%
  \BibitemOpen
  \bibfield  {author} {\bibinfo {author} {\bibfnamefont {A.}~\bibnamefont
  {Eggleton}}\ and\ \bibinfo {author} {\bibfnamefont {F.}~\bibnamefont
  {Tompkins}},\ }\href@noop {} {\bibfield  {journal} {\bibinfo  {journal}
  {Transactions of the Faraday Society}\ }\textbf {\bibinfo {volume} {48}},\
  \bibinfo {pages} {738} (\bibinfo {year} {1952})}\BibitemShut {NoStop}%
\bibitem [{\citenamefont {Liepmann}(1961)}]{liepmann_1961}%
  \BibitemOpen
  \bibfield  {author} {\bibinfo {author} {\bibfnamefont {H.~W.}\ \bibnamefont
  {Liepmann}},\ }\href {\doibase 10.1017/S002211206100007X} {\bibfield
  {journal} {\bibinfo  {journal} {Journal of Fluid Mechanics}\ }\textbf
  {\bibinfo {volume} {10}},\ \bibinfo {pages} {65–79} (\bibinfo {year}
  {1961})}\BibitemShut {NoStop}%
\bibitem [{\citenamefont {Merritt}\ \emph {et~al.}(2014)\citenamefont
  {Merritt}, \citenamefont {Moser}, \citenamefont {Hsu}, \citenamefont {Adams},
  \citenamefont {Dunn}, \citenamefont {Miguel~Holgado},\ and\ \citenamefont
  {Gilmore}}]{collidingJets}%
  \BibitemOpen
  \bibfield  {author} {\bibinfo {author} {\bibfnamefont {E.~C.}\ \bibnamefont
  {Merritt}}, \bibinfo {author} {\bibfnamefont {A.~L.}\ \bibnamefont {Moser}},
  \bibinfo {author} {\bibfnamefont {S.~C.}\ \bibnamefont {Hsu}}, \bibinfo
  {author} {\bibfnamefont {C.~S.}\ \bibnamefont {Adams}}, \bibinfo {author}
  {\bibfnamefont {J.~P.}\ \bibnamefont {Dunn}}, \bibinfo {author}
  {\bibfnamefont {A.}~\bibnamefont {Miguel~Holgado}}, \ and\ \bibinfo {author}
  {\bibfnamefont {M.~A.}\ \bibnamefont {Gilmore}},\ }\href {\doibase
  10.1063/1.4872323} {\bibfield  {journal} {\bibinfo  {journal} {Physics of
  Plasmas}\ }\textbf {\bibinfo {volume} {21}},\ \bibinfo {pages} {055703}
  (\bibinfo {year} {2014})},\ \Eprint
  {http://arxiv.org/abs/https://doi.org/10.1063/1.4872323}
  {https://doi.org/10.1063/1.4872323} \BibitemShut {NoStop}%
\bibitem [{\citenamefont {Underwood}(2019)}]{underwood2019hydromagnetic}%
  \BibitemOpen
  \bibfield  {author} {\bibinfo {author} {\bibfnamefont {T.~C.}\ \bibnamefont
  {Underwood}},\ }\href@noop {} {\emph {\bibinfo {title} {Hydromagnetic
  stability and collisional properties of current-driven plasma jets}}}\
  (\bibinfo  {publisher} {Stanford University},\ \bibinfo {year}
  {2019})\BibitemShut {NoStop}%
\end{thebibliography}%

\end{document}